\documentclass[pra,twocolumn,showpacs]{revtex4}
\usepackage{epsfig,psfrag}
\usepackage{amsmath,amsfonts,bm}
\usepackage{mathrsfs}
\usepackage{revsymb}
\usepackage{exscale}

\usepackage[T1]{fontenc}
\usepackage{ae,aecompl} 

\begin{document}

\title{How to measure  the wavefunction  absolute squared of a
         moving particle by using mirrors}

\author{Volker Hannstein}
\author{Gerhard C.~Hegerfeldt}

\affiliation{Institut f\"ur Theoretische Physik, Universit\"at G\"ottingen,
  Tammannstr.~1, 37077 G\"ottingen, Germany}

\pacs{03.65.Wj, 03.75.Be, 32.80.Lg}

\newcommand{\nfrac}[2]{\genfrac{}{}{0pt}{}{#1}{#2}} 
\renewcommand{\vec}[1]{\bm{#1}}          
\renewcommand{\t}[1]{\text{#1}}          
\newcommand{\e}[1]{\mathrm{e}^{#1}}          
\renewcommand{\Re}{\mathrm{Re}\,}            
\renewcommand{\Im}{\mathrm{Im}\,}            
\renewcommand{\i}{\mathrm{i}}                
\renewcommand{\d}[1]{\t{d}#1\,}              

\begin{abstract}
We consider a slow particle with  wavefunction
$\psi_t(\vec{x})$, moving freely in some direction. A mirror 
is briefly switched on around a time $T$ and its position
is scanned. It is shown that the measured reflection probability then
allows the determination of $|\psi_T(\vec{x})|^2$. Experimentally
available atomic mirrors should make this method applicable to the
center-of-mass wavefunction of atoms with velocities in the cm/s range.
\end{abstract}

\maketitle

Measuring the absolute value of a wavefunction is part of the more
general quest to reconstruct a  quantum state from a series of
repeated measurements on a system \cite{PaRe:04,Le:97,BaLeScSc:96,LeMeKiMoIt:96,Mo:02,ScHaRiGuLa:03,KuPfMl:97,BaRaSu:03,KoPr:97,DhKoRuHaRo:97,RuDhKoRoSm:99,PrRuDhKoSm:99,RuKoDhRoGu:99,KoRoRuSmPr:00,KaQaZu:03}.
Physically, the absolute square, $|\psi_t(\vec{x})|^2$, of the
wavefunction of a particle  gives the probability density for finding the
particle at the position $\vec{x}$, at time $t$. Hence, to determine
it experimentally for an unknown wavefunction one just would  have to perform
a series of repeated position measurements. In general, the difficulty
with this  procedure is  the required precision since it  must be
much better than the spatial extent of the wavefunction. An additional problem
occurs for slow laser cooled atoms since their wavefunction can be
influenced by a measurement. Here we are mainly interested
in the center-of-mass (CM) motion of such slow  laser-cooled atoms
with wavefunction widths in the region of a few micrometers and velocities in the cm/s
range. Such states can be  obtained by letting atoms escape from a trap
\cite{SzGuArDa:96}.  

An elegant position measurement technique \cite{SaDaAsMeCo:87}, which
would have the necessary resolution, places the atom in a
strong gradient of a magnetic or light field. In this field the internal
energy levels vary rapidly with position.  By
sending a light beam at one of the atomic resonance frequencies one
then detects the atoms which are located where the resonance
occurs. In this way, sub-wavelength resolution can be
achieved. However, when the atom moves through the field its CM
wavefunction will, except under special circumstances, be affected by the
interaction and will not remain the original free CM
wavefunction. Hence one might not obtain $|\psi_t(\vec{x})|^2$ but some
other distribution.   

In this paper we propose an alternative method for determining
$|\psi_t(\vec{x})|^2$ for very slow free particles, in particular for 
laser cooled 
atoms.  This  method  uses reflection by a  mirror which is 
switched on for a brief time period $\tau$ and whose position is scanned. The
reflection probability, which  can be  determined
sufficiently far away from the mirror by fluorescence
techniques,  will be shown to allow the  
determination of $|\psi_t(\vec{x})|^2$.  

\begin{figure}[bt]
\epsfig{file=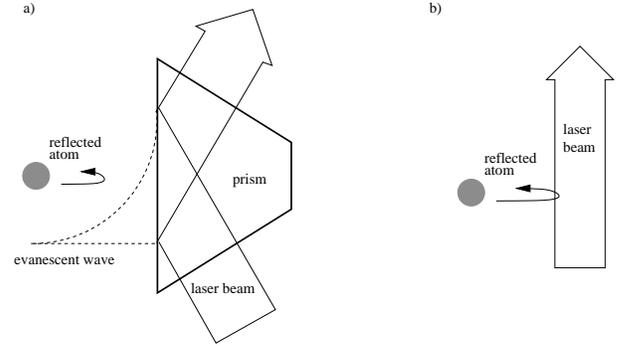,width=8cm}
\caption{\label{fig:mirror} {Schematic scetch of an evanescent wave
    mirror. The atoms are reflected by the potential created by the
    evanescent wave which occurs on the surface of the prism. }} 
\end{figure}

A suitable mirror is an evanescent wave (EW) mirror 
\cite{CoHi:82,BaLeOvSi:87,BaLeOvSi:88,KaWeCh:90},
depicted in fig.~\ref{fig:mirror}. A laser beam enters 
a prism and is totally reflected at the surface, creating an
evanescent wave outside the prism. Due to the intensity gradient an
atom positioned there experiences  a steep exponential potential
which is repulsive if 
the laser frequency is above the atomic resonance frequency (blue
detuning). For a strong laser, a slow atom will be reflected without
reaching the surface. Such a mirror
is  easily switched on and off by an acousto-optic
modulator. Switching times of .5 $\mu$s or less can be
implemented. Reflection takes place within a region in the order of
$0.1{\mu m}$. For our
purposes  the experimental setup could be  
similar to the  temporal two-slit experiment of Ref. \cite{SzGuArDa:96}.

To simplify the analysis we replace the exponential potential of the
mirror by a high square potential barrier. Letting  
its height go to infinity and its thickness go to zero  the mirror is
modeled by a reflecting plane. Possible adsorption on the prism of the
mirror is disregarded. This modeling simplifies the mathematics but keeps the
essential physics. It will be shown that for small $\tau$ the
reflection probability scales as 
$\sqrt{\tau}$, multiplied by a factor depending on the absolute square
of the wavefunction at the position of the mirror. It is also shown
that for larger $\tau$ there is a region of linear dependence.

\begin{figure}[bt]
\epsfig{file=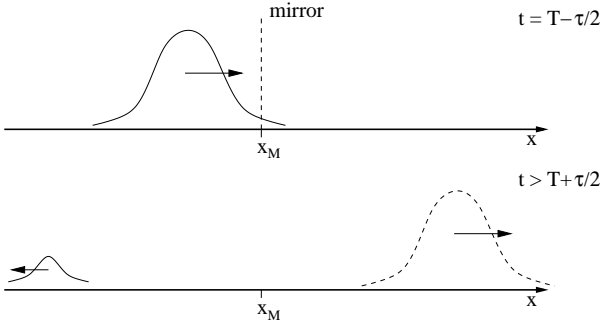,width=8cm}
\caption{\label{1d} {Wave packet $\psi_t(x)$ coming in from the left
    (solid line). 
    Mirror (dashed) at $x_\t{M}$ switched on at $T-\tau/2$ and
    switched off at $T+\tau/2$.
    Partially reflected wave (dotted). Measured reflection probability
    allows the determination of $|\psi_T(x_\t{M})|^2$.  }} 
\end{figure}

First we consider the one-dimensional case, with a free wavefunction
 coming  from the left and traveling to the
right (see fig.~\ref{1d}), which is denoted by $\psi_t(x)$ if there is
no mirror present. When the mirror is  active from $T-\tau/2$ to 
$T+\tau/2$ the wavefunction evolves freely until time $T-\tau/2$,
experiences the mirror from $T-\tau/2$ to $T+\tau/2$ and then again evolves
freely. The mirror action causes partial reflection, and the
reflected part can be calculated either by an eigenfunction expansion, 
or with Green's functions as in the following.

The mirror is first positioned at $x_M=0$. The free time development
operator $U^\t{F}(t,0)$ is given in terms of the free Green's function
$G_t^\t{F}(x)$ by 
\begin{multline}
\langle x|U^\t{F}(t,0)|x'\rangle \\ \equiv G_t^\t{F}(x-x') =
\sqrt{m/2\pi\i\hbar t}\,\e{-m (x-x')^2/2\i\hbar t}. 
\end{multline}
During the time the mirror is active the time development
operator $U^\pm(t,0)$ on the
 right $(+)$ and left $(-)$ half axis, respectively, is given in terms of the
 Green's function $G_t^\pm(x,x')$ by
\begin{multline}\nonumber
\langle x|U^\pm(t,0)|x'\rangle \equiv G_t^\pm(x,x') \\ = \theta_\pm(x)
(G_t^\t{F}(x-x')-G_t^\t{F}(x+x'))\theta_\pm(x') 
\end{multline} 
where $\theta_\pm(x)\equiv\theta(\pm x)$ and $\theta(x)$ is the
Heaviside function. 
Let $\phi^\pm_t$ denote the wavefunction
on the right and left half axis, respectively, when  the mirror is
active. With $\psi_{T-\tau/2}$ the (free) wavefunction on the
whole axis at time ${T-\tau/2}$, one  has $\phi^\pm_{T+\tau/2} =
U^\pm(\tau,0)(\theta_\pm\psi_{T-\tau/2})$, i.e. 
\begin{multline}
\label{eq:psipm}
\phi^\pm_{T+\tau/2}(x) = \theta_\pm(x)
(G_\tau^\t{F}\ast(\theta_\pm\psi_{T-\tau/2}))(x)\\ - \theta_\pm(x) 
(G_\tau^\t{F}\ast \hat{\pi}(\theta_\pm\psi_{T-\tau/2}))(x),
\end{multline}
where $\ast$ denotes convolution and $\hat{\pi}$ reflection: $x
\rightarrow -x$. The complete wavefunction  at the switch-off
time is $\phi_{T+\tau/2}(x)= \phi^+_{T+\tau/2}(x) + \phi^-_{T+\tau/2}(x)$,
which then evolves freely at later times. The reflected part is 
given by the Fourier transform 
$\widetilde{\phi}_{T+\tau/2}(k)$ for $k<0$ and the reflection
probability, denoted by $N_\t{refl}(\tau,T;x_M)$, is given by its  norm
squared, i.e. for $x_M =0$ 
\begin{equation}
\label{eq:reflnorm}
N_\t{refl}(\tau,T;0) = \int_{-\infty}^0\d{k}|\widetilde{\phi}_{T+\tau/2}(k)|^2.
\end{equation}
Fourier transforming eq.~\eqref{eq:psipm} one obtains after some
calculation
\begin{multline}
\label{eq:psireflb}
\widetilde{\phi}_{T+\tau/2}(k) 
= \frac{1}{2}\Big(\widetilde{\psi}_{T-\tau/2}(k)-\widetilde{\psi}_{T-\tau/2}(-k)\Big)\e{-\i\alpha 
  k^2} \\ + \frac{1}{2\pi^2}\mathscr{P}\!\!\int_{-\infty}^\infty \d{k'}\Big[
\frac{1}{k'-k} + \frac{1}{k'+k}\Big]\e{-\i\alpha
  k'^2}  \\ \times \mathscr{P}\!\int_{-\infty}^\infty \d{k''}
\frac{\widetilde{\psi}_{T-\tau/2}(k'')}{k'-k''} 
\end{multline}
where $\mathscr{P}$ denotes Cauchy's principal value and 
\begin{equation}
\alpha\equiv\hbar \tau/2m~.
\end{equation}
For $\tau \to 0$  the right hand side of eq.~\eqref{eq:psireflb} has to
converge to $\widetilde{\psi}_{T}(k)$ which shows that there must
be compensations between the first and second term. As a consequence, 
the behavior for small $\tau$ (i.e. small $\alpha$) is not
easily extracted from this equation. We therefore use the
skew-reciprocity of the Hilbert transform \cite{Hilbert} to write
\begin{multline}
\frac{1}{2}\Big(\widetilde{\psi}_{T-\tau/2}(k)-
\widetilde{\psi}_{T-\tau/2}(-k)\Big)\e{-\i\alpha  k^2} = \\ 
 \widetilde{\psi}_{T-\tau/2}(k)\e{-\i\alpha k^2} 
+ \frac{1}{2\pi^2}\mathscr{P}\int_{-\infty}^\infty \d{k'}\Big[
\frac{1}{k'-k} + \frac{1}{k'+k}\Big]\\ \times  
\mathscr{P}\int_{-\infty}^\infty \d{k''} 
\frac{\widetilde{\psi}_{T-\tau/2}(k'')}{k'-k''}\e{-\i\alpha
  k''^2}  
\end{multline}
and then use
the anti-symmetry in $k'$ of the square bracket in
eq.~\eqref{eq:psireflb} to obtain, with the substitution $\kappa=
k'/\sqrt{\alpha}$ and 
$\widetilde{\psi}_{T-\tau/2}(k)=\e{\i\alpha k^2/2}\widetilde{\psi}_T(k)$,
\begin{multline}
\label{eq:psireflfree}
\widetilde{\phi}_{T+\tau/2}(k) 
= \widetilde{\psi}_T(k)\e{-\i\alpha k^2/2} \\ +
  \frac{\sqrt{\alpha}}{2\pi^2}\mathscr{P}\int_{-\infty}^\infty
  \d{\kappa}\Big[ 
\frac{1}{\kappa-\sqrt{\alpha}k} +
\frac{1}{\kappa+\sqrt{\alpha}k}\Big]  
\\ \times \int_{-\infty}^\infty \d{k''} 2\kappa\frac{\e{-\i \kappa^2}
   - \e{-\i\alpha k''^2}}{\kappa^2-\alpha k''^2}\e{\i\alpha
  k''^2/2}\widetilde{\psi}_T(k''). 
\end{multline}
The last, now non-singular, integrand can be expanded in
$\sqrt\alpha$, i.e. $\sqrt\alpha k \ll 1$ for
typical $k$ values of $\tilde\psi$,  which in lowest
order yields for eq.~\eqref{eq:psireflfree}
\begin{multline}
\label{eq:psireflfirstord}
\widetilde{\phi}_{T+\tau/2}(k) = \widetilde{\psi}_T(k)\e{-\i\alpha k^2/2}
\\ + \sqrt{\alpha}\frac{\sqrt{2\pi}}{\pi^2}\psi_T(0)
\mathscr{P}\int_{-\infty}^\infty  
\d{\kappa}\frac{1}{\kappa-\sqrt{\alpha}k}\frac{\e{-\i \kappa^2}-1}{\kappa}
\end{multline}
where $\int
\d{k''}\widetilde{\psi}_T(k'')=\sqrt{2\pi}\,\psi_T(0) $
has been used. For $k<0$ one has $\widetilde{\psi}_T(k)=0$ since the free 
particle is coming in from the left. The
reflected wavefunction has a $\sqrt{\alpha}$ in front and from this one
might, erroneously, expect that the reflection probability goes as
$\alpha$, i.e. as $\tau$. However, calculating the reflection
probability  with 
Eqs.~\eqref{eq:reflnorm} and (\ref{eq:psireflfirstord}), while
using  the symmetry of the appearing integrands and employing
footnote \cite{Hilbert}, the  substitution $k\rightarrow k/\sqrt\alpha$ gives
an additional $1/\sqrt{\alpha}$ and one obtains, with the mirror 
located at $x_\t{M}=x$,
\begin{equation}
\label{eq:NtauOrd1}
N_\t{refl}(\tau,T;x_\t{M}=x) = 2\sqrt{\frac{\hbar\tau}{\pi
    m}}|\psi_T(x)|^2 
\end{equation}  
in lowest order in $\sqrt{\tau}$, with the mirror active from
${T-\tau/2}$ to $T+\tau/2$.  
Since $\int \d{x} |\psi_T|^2=1$, eq.~\eqref{eq:NtauOrd1} implies 
\begin{equation}\label{Nnormed}
|\psi_T(x)|^2 = N_\t{refl}(\tau,T;x_\t{M}=x)/\int \d{x'}
N_\t{refl}(\tau,T;x'). 
\end{equation}
Going to higher orders in the expansion of the last integral in
eq.~\eqref{eq:psireflfree} one obtains after some calculation, again
using  the symmetry of the appearing integrands and employing
footnote \cite{Hilbert},
\begin{multline}
\label{eq:NtauOrd3}
N_\t{refl}(\tau,T;x_\t{M}=x) =  
2\sqrt{\frac{\hbar \tau}{\pi m}} \bigg[ |\psi_T(x)|^2  \\
- \frac{\hbar\tau}{6m}\Re
\big(\overline{\psi_T(x)}\psi_T''(x)\big)   
+\mathcal{O}(\tau^2)\bigg]. 
\end{multline}
Thus eq.~\eqref{eq:NtauOrd1} is expected to be a good
approximation if  $\tau$ satisfies
$\tau \ll 6m|\psi_T(x_\t{M})|/\hbar|\psi_T''(x_\t{M})|$.
This shows that if the curvature of the wavefunction is large one
needs short 
switch-on durations of the mirror for eq.~\eqref{eq:NtauOrd1} to be a
good approximation. Equation \eqref{Nnormed}, on the other hand, turns out
to be valid for a much wider range of mirror pulses $\tau$.

\begin{figure}[tb]
  \epsfig{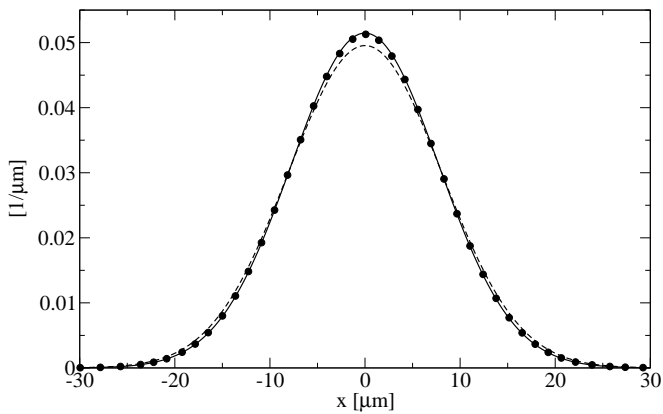}
  \caption{\label{fig:tempslit_x_exp} Slow
    ($v_0=25~$cm/s) initially minimal Gaussian
    wave packet for Cs atoms prepared at time $t=0$ with mean
  position $x_0=-12~$mm, initial width $\Delta_x=
  1.6~\mu$m. Mirror switched on at
  $T=50~$ms. Plotted:
    $|\psi_T(x)|^2$  
      (solid line) and r.h.s of eq.~(\ref{Nnormed}) for
      $\tau=10~\mu$s
      (dots) and $\tau=30~\mu$s (dashed line). }
\end{figure}

It will now be shown that there is  a linear regime  for larger
$\tau$ and that eq.~\eqref{Nnormed} also holds in this regime. 
By means of Eqs.~\eqref{eq:reflnorm} and
(\ref{eq:psireflfree}) one can write $N_\t{refl}$ in the form
\begin{equation}
N_\t{refl}(\tau,T;0) = \int \d{k_1} \d{k_2} \overline{\widetilde\psi_T(k_1)}I(k_1,k_2){\widetilde\psi_T(k_2)}
\end{equation}
where a lengthy calculation gives for the integral kernel 
\begin{multline}
I(k_1,k_2) =\frac {\i}{2\pi \left( k_2^{2}-  k_1^{2} \right) }
 \\ \times \Bigl\{\e{-\i\alpha ( k_2^2-  k_1^2 )/2}\bigl[ k_1 \t{erf}
  (\sqrt{\i\alpha} k_1) +  k_2\,  \t{erf} (\sqrt{-\i\alpha} k_2
  )\bigr]  \\   - \e{\i\alpha ( k_2^2-  k_1^2 )/2}\bigl[ k_2 \,  \t{erf} (
  \sqrt{\i\alpha} k_2) +  k_1\,  \t{erf}(\sqrt{-\i\alpha} k_1)\bigr] 
  \Bigr\}.
 \label{kernel}
\end{multline}
Expanding this for small $\alpha$, i.e. $\sqrt\alpha k \ll 1$ for
typical $k$ values of $\tilde\psi$, one recovers Eqs.
(\ref{eq:NtauOrd1}) and (\ref{eq:NtauOrd3}). On the other hand, 
when $\sqrt\alpha k \gg 1$  one can use the asymptotic expansion of
the error function, $\t{erf}\, z = 1 + \pi ^{-1/2}\,e^{-z2}
\{-1/z + 1/2z^3 + \mathcal{O}(z^{-5})  \} $. With this one obtains in
lowest order
\begin{equation} 
\label{kernel0}
I(k_1,k_2) = \frac{\alpha}{2\pi}(k_1 + k_2) \frac{\sin \alpha(k_2^2 -
  k_1^2)/2}{\alpha(k_2^2 - k_1^2)/2}~.
\end{equation}
The contributions from the next order cancel and  $z^{-3}$ 
yields an $\alpha^{-3/2}$ term.
Provided $\alpha k_0 \Delta_k$ is
small, with $\hbar k_0$ the mean momentum and $\hbar \Delta_k$ the momentum
width, the last fraction in eq.~\eqref{kernel0} can be approximated by
1 and $k_1$ and $k_2$ by $k_0$. Then, in this order, 
$N_\t{refl}(\tau,T;x_\t{M}=x) = 2 |\psi_T(x)|^2 k_0\, \alpha \equiv
|\psi_T(x)|^2 v_0 \tau$, which yields
eq.~\eqref{Nnormed}. Under the same assumptions the next nonvanishing
correction is small. 
 Hence in this regime the reflection probability is linear in $\tau$ and
 eq.~\eqref{Nnormed} is a good approximation \cite{infty}. For the wavefunction
 considered in fig.~\ref{fig:tempslit_x_exp}
 one has $\sqrt\alpha k = 25 \gg 1$ and $\alpha k_0 \Delta_k \leq .8$ which
 is sufficient in this case.

\begin{figure}[bt]
\epsfig{file=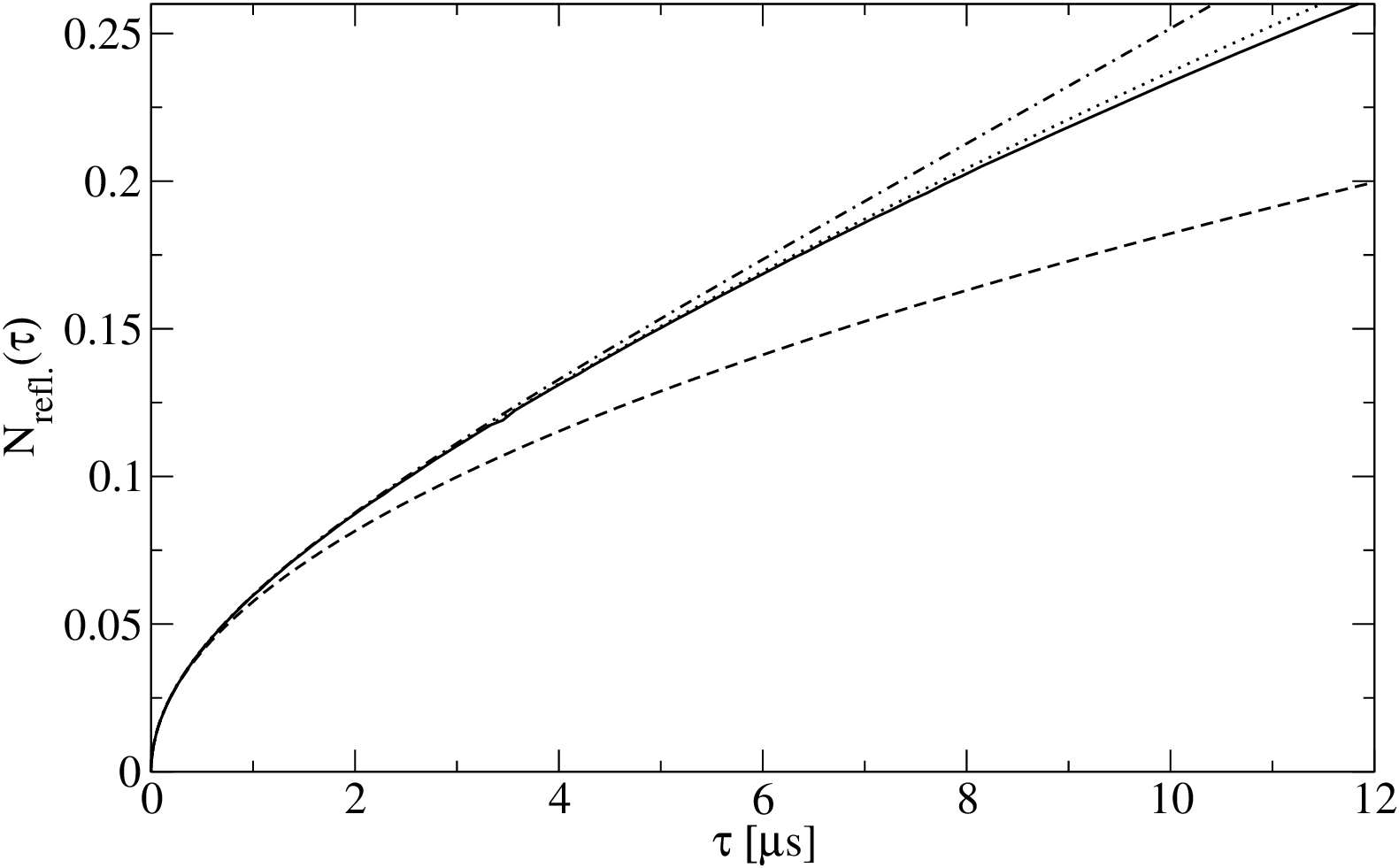,width=6.5cm}
\epsfig{file=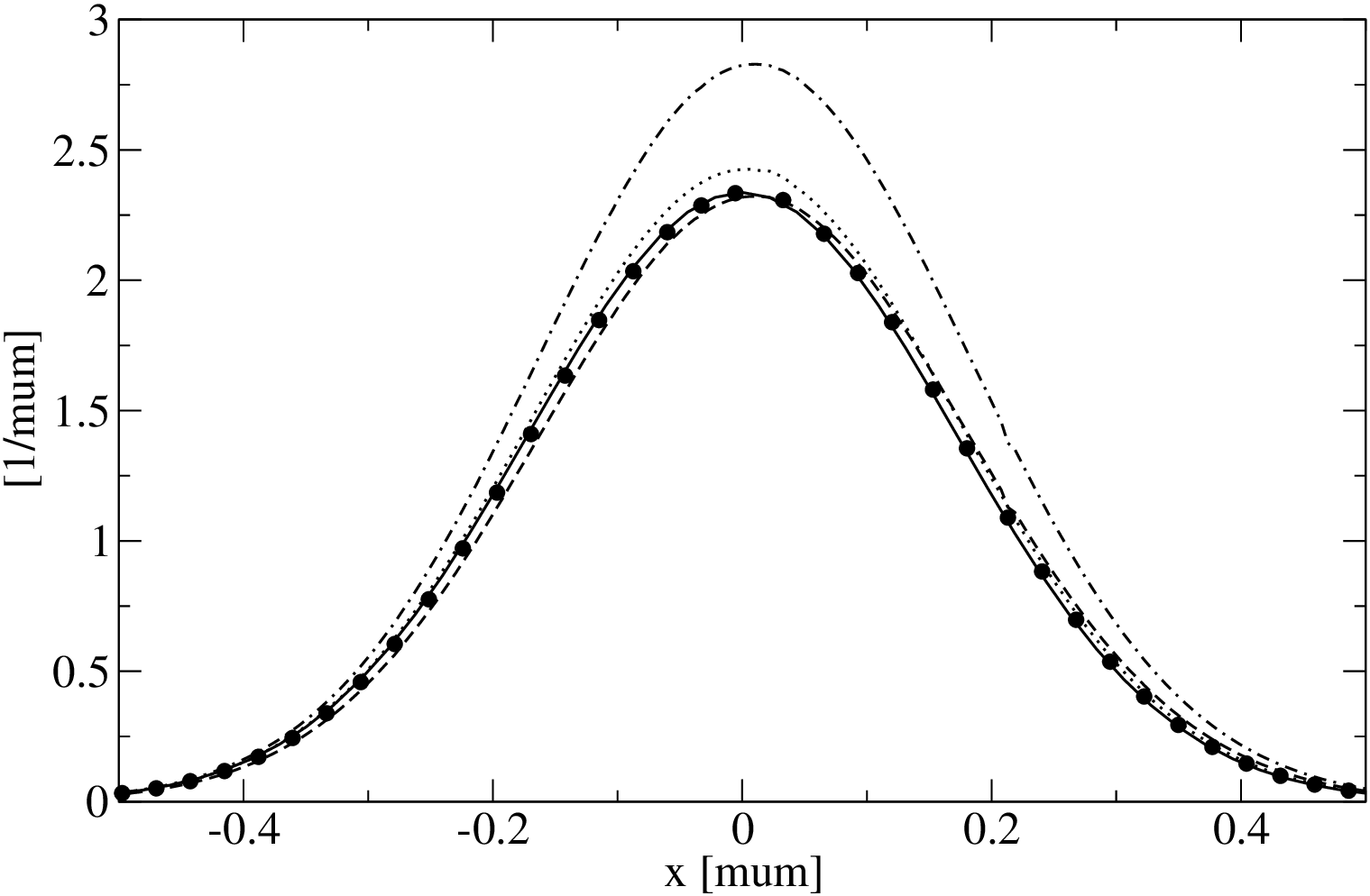,width=6.5cm}
\caption{\label{fig:temp_tau_slow} { Reflection probability
    vs.   pulse duration $\tau$ for a slow
    ($v_0=1.1\,\t{cm}/\t{s}$) initially minimal Gaussian
    wave packet for Cs atoms prepared at time $t=0$ with mean
  position $x_0=-0.66\,\mu\t{m}$, initial width $\Delta_x=
  0.1\,\mu$m. Mirror switched on at
  $T=60\,\mu$s. Solid line: numerical result.
Dashed line: lowest order (eq.~(\ref{eq:NtauOrd1})).
Dashed-dotted line: up to second order. Dotted line: up to third order.}}
\caption{\label{fig:temp_x_slow} { Wavefunction and $T$ as in
    fig.~\ref{fig:temp_tau_slow}.  Solid line: 
    $|\psi_T(x)|^2$. Dashed-dotted line: $N_\t{refl}(\tau,T;x_\t{M}=x)/2\sqrt{\hbar
      \tau/\pi m}$ for $\tau = 1 ~\mu$s.  Dotted: the same for
    $\tau = 5 ~\mu$s. Fat dots: r.h.s of eq.~(\ref{Nnormed}) for $\tau
    = 1 ~\mu$s.  Dashed: r.h.s of eq.~(\ref{Nnormed}) for $\tau
    = 5 ~\mu$s.}} 
\end{figure}

In fig.~\ref{fig:temp_tau_slow}, a numerically calculated reflection
probability $N_\t{refl}(\tau,T;x_\t{M}=0)$ as a function of $\tau$ is 
 compared with various analytic
 approximations. For the chosen example of a Gaussian wavefunction of
 a Cs atom  with  velocity 
 $v_0=1.1~$cm/s it is seen that  eq.~\eqref{eq:NtauOrd1}
 agrees well with the  exact result for $\tau\leq 1.5~ \mu$s. For higher
 velocities  shorter pulse durations $\tau$ are
 needed in eq.~\eqref{eq:NtauOrd1}. If these are too short for
 available mirrors  one may  either fit the higher 
 order result in eq.~\eqref{eq:NtauOrd3} to  experimentally
 obtained reflection probabilities  for different   pulse
 durations $\tau$ or, better, use  eq.~\eqref{Nnormed} in the regime
 linear in $\tau$.

 In fig.~\ref{fig:temp_x_slow}  we  consider the same wavefunction as in 
fig.~\ref{fig:temp_tau_slow}. The solid line is $|\psi_{T}(x)|^2$. This
is compared with the numerically obtained reflection probability, 
$N_\t{refl}(\tau,T;x_\t{M}=x)$, for different mirror positions $x$,  divided by
$2\sqrt{\hbar \tau/\pi m}$  in view of eq.~\eqref{eq:NtauOrd1}.  For
$\tau = 1~ \mu$s (dashed-dotted line) the 
agreement is good, but not quite so good for $\tau = 5~\mu$s (dotted
line), as expected from fig.~\ref{fig:temp_tau_slow}. 
Also in fig.~\ref{fig:temp_x_slow} is plotted the
right hand side  of eq.~\eqref{Nnormed}  for $\tau = 1 ~\mu$s and
$\tau = 5~\mu$s. For $\tau = 1 ~\mu$s (fat dots) it is practically indistinguishable from
$|\psi_T(x)|^2$ and now also for $\tau = 5~\mu$s (dashed line)  the
agreement is excellent. The slight shift is due to the next order
term in eq.~\eqref{eq:NtauOrd3}. 

A test of eq.~\eqref{Nnormed} in the linear regime  is displayed 
 in fig.~\ref{fig:tempslit_x_exp} for  a slow ($v_0=25~$cm/s)
 wavefunction  for which the lowest order   
expressions in Eqs.~\eqref{eq:NtauOrd1} and \eqref{eq:psireflfirstord}
are valid only for  unrealistically short pulse durations of less than
$0.01~\mu$s. However,  even for pulse durations of  up to $\tau = 30~
\mu$s the  right hand side of eq.~\eqref{Nnormed} agrees extremely well with
$|\psi_T(x)|^2$. This is remarkable since during $30~ \mu$s the
wavefunction moves by $7.5~ \mu$m, a third of its width at time $T$, and
the agreement is attributed to interference.

In the three-dimensional case the mirror is first assumed to be positioned
at $x_M$ and parallel to 
the $y-z$ plane.  The wave packet is assumed to be incident from the
negative $x$ direction, i.e.~$\widetilde{\psi}_t(\vec{k})$ is nonzero for
$k_x>0$ only. The analysis is analogous to the one-dimensional case
and one obtains
\begin{multline}\label{three}
\int\d{y}\int\d{z}|\psi_T(x_\t{M},y,z)|^2 \\ =
N_\t{refl}(\tau,T;x_\t{M}) / \int \d{x'} N_\t{refl}(\tau,T;x') 
\end{multline}
both in the square-root and linear regime.
By this experimental setup one therefore obtains the 
position space density for the wavefunction integrated over the mirror
plane \cite{T}. By rotating the mirror by different angles around the point 
$\vec{x}_\t{M}$ one can also obtain the integrated atomic density in
these rotated planes. In addition one may also vary the point
$\vec{x}_\t{M}$. This is similar to the situation  encountered in
computer tomography, and in this way one can recover
$|\psi_T(\vec{x})|^2$ from the experimental data by standard techniques.
The presence of a perpendicular gravitational field does not change
the conclusions 
provided the pulse duration is sufficiently short. Indeed, until the
mirror is switched on the free evolution is modified by the gravitational
potential, but during the short pulse time its effect is negligible
and the reflection probability is similar as before.

In summary, we have proposed to measure the absolute square of the
 wavefunction of a slowly moving particle by using
reflection from a 
pulsed  mirror. It has been shown that for short pulse duration
$\tau$  the reflection probability scales as $\sqrt \tau$ and for
somewhat longer duration as $\tau$. By measuring the reflection
probability and scanning the mirror position one can determine
$|\psi_t(\vec{x})|^2$. With presently available atomic mirrors it
should be possible to apply this method to the center-of-mass
wavefunction of atoms with velocities in the cm/s range. This
technique provides a sub-wavelength resolution and has the advantage
of avoiding, at least in principle, the problem  
of a possible disturbance of the wavefunction during the measurement.


\begin{thebibliography}{26}
\expandafter\ifx\csname natexlab\endcsname\relax\def\natexlab#1{#1}\fi
\expandafter\ifx\csname bibnamefont\endcsname\relax
  \def\bibnamefont#1{#1}\fi
\expandafter\ifx\csname bibfnamefont\endcsname\relax
  \def\bibfnamefont#1{#1}\fi
\expandafter\ifx\csname citenamefont\endcsname\relax
  \def\citenamefont#1{#1}\fi
\expandafter\ifx\csname url\endcsname\relax
  \def\url#1{\texttt{#1}}\fi
\expandafter\ifx\csname urlprefix\endcsname\relax\def\urlprefix{URL }\fi
\providecommand{\bibinfo}[2]{#2}
\providecommand{\eprint}[2][]{\url{#2}}

\bibitem[{\citenamefont{Paris and \v{R}eh\'a\v{c}ek}(2004)}]{PaRe:04}
\bibinfo{editor}{\bibfnamefont{M.}~\bibnamefont{Paris}} \bibnamefont{and}
  \bibinfo{editor}{\bibfnamefont{J.}~\bibnamefont{\v{R}eh\'a\v{c}ek}}, eds.,
  \emph{\bibinfo{title}{Quantum State Estimation}}, vol. \bibinfo{volume}{649}
  of \emph{\bibinfo{series}{Lecture Notes in Physics}}
  (\bibinfo{publisher}{Springer}, \bibinfo{year}{2004}).

\bibitem[{\citenamefont{Leonhardt}(1997)}]{Le:97}
\bibinfo{author}{\bibfnamefont{U.}~\bibnamefont{Leonhardt}},
  \emph{\bibinfo{title}{Measuring the quantum state of light}}, Cambridge
  studies in modern physics (\bibinfo{publisher}{Cambridge Univ. Press},
  \bibinfo{year}{1997}).

\bibitem[{\citenamefont{Bardoff et~al.}(1996)\citenamefont{Bardoff, Leichtle,
  Schrade, and Schleich}}]{BaLeScSc:96}
\bibinfo{author}{\bibfnamefont{P.~J.} \bibnamefont{Bardoff}},
  \bibinfo{author}{\bibfnamefont{C.}~\bibnamefont{Leichtle}},
  \bibinfo{author}{\bibfnamefont{G.}~\bibnamefont{Schrade}}, \bibnamefont{and}
  \bibinfo{author}{\bibfnamefont{W.~P.} \bibnamefont{Schleich}},
  \bibinfo{journal}{Phys. Rev. Lett.} \textbf{\bibinfo{volume}{77}},
  \bibinfo{pages}{2198} (\bibinfo{year}{1996}).

\bibitem[{\citenamefont{Leibfried et~al.}(1996)\citenamefont{Leibfried,
  Meekhof, King, Monroe, Itano, and Wineland}}]{LeMeKiMoIt:96}
\bibinfo{author}{\bibfnamefont{D.}~\bibnamefont{Leibfried}},
  \bibinfo{author}{\bibfnamefont{D.~M.} \bibnamefont{Meekhof}},
  \bibinfo{author}{\bibfnamefont{B.~E.} \bibnamefont{King}},
  \bibinfo{author}{\bibfnamefont{C.}~\bibnamefont{Monroe}},
  \bibinfo{author}{\bibfnamefont{W.~M.} \bibnamefont{Itano}}, \bibnamefont{and}
  \bibinfo{author}{\bibfnamefont{D.~J.} \bibnamefont{Wineland}},
  \bibinfo{journal}{Phys. Rev. Lett.} \textbf{\bibinfo{volume}{77}},
  \bibinfo{pages}{4281} (\bibinfo{year}{1996}).

\bibitem[{\citenamefont{Monroe}(2002)}]{Mo:02}
\bibinfo{author}{\bibfnamefont{C.}~\bibnamefont{Monroe}},
  \bibinfo{journal}{Nature} \textbf{\bibinfo{volume}{416}},
  \bibinfo{pages}{238} (\bibinfo{year}{2002}).

\bibitem[{\citenamefont{Schmidt-Kaler et~al.}(2003)\citenamefont{Schmidt-Kaler,
  H\"affner, Riebe, Gulde, Lancaster, Deuschle, Becher, Roos, Eschner, and
  Blatt}}]{ScHaRiGuLa:03}
\bibinfo{author}{\bibfnamefont{F.}~\bibnamefont{Schmidt-Kaler}},
  \bibinfo{author}{\bibfnamefont{H.}~\bibnamefont{H\"affner}},
  \bibinfo{author}{\bibfnamefont{M.}~\bibnamefont{Riebe}},
  \bibinfo{author}{\bibfnamefont{S.}~\bibnamefont{Gulde}},
  \bibinfo{author}{\bibfnamefont{G.~P.~T.} \bibnamefont{Lancaster}},
  \bibinfo{author}{\bibfnamefont{T.}~\bibnamefont{Deuschle}},
  \bibinfo{author}{\bibfnamefont{C.}~\bibnamefont{Becher}},
  \bibinfo{author}{\bibfnamefont{C.~F.} \bibnamefont{Roos}},
  \bibinfo{author}{\bibfnamefont{J.}~\bibnamefont{Eschner}}, \bibnamefont{and}
  \bibinfo{author}{\bibfnamefont{R.}~\bibnamefont{Blatt}},
  \bibinfo{journal}{Nature} \textbf{\bibinfo{volume}{422}},
  \bibinfo{pages}{408} (\bibinfo{year}{2003}).

\bibitem[{\citenamefont{Kurtsiefer et~al.}(1997)\citenamefont{Kurtsiefer, Pfau,
  and Mlynek}}]{KuPfMl:97}
\bibinfo{author}{\bibfnamefont{C.}~\bibnamefont{Kurtsiefer}},
  \bibinfo{author}{\bibfnamefont{T.}~\bibnamefont{Pfau}}, \bibnamefont{and}
  \bibinfo{author}{\bibfnamefont{J.}~\bibnamefont{Mlynek}},
  \bibinfo{journal}{Nature} \textbf{\bibinfo{volume}{386}},
  \bibinfo{pages}{150} (\bibinfo{year}{1997}).

\bibitem[{\citenamefont{Baron et~al.}(2003)\citenamefont{Baron, Rauch, and
  Suda}}]{BaRaSu:03}
\bibinfo{author}{\bibfnamefont{M.}~\bibnamefont{Baron}},
  \bibinfo{author}{\bibfnamefont{H.}~\bibnamefont{Rauch}}, \bibnamefont{and}
  \bibinfo{author}{\bibfnamefont{M.}~\bibnamefont{Suda}}, \bibinfo{journal}{J.
  Opt. B: Quantum Semiclass. Opt.} \textbf{\bibinfo{volume}{5}},
  \bibinfo{pages}{S241} (\bibinfo{year}{2003}).

\bibitem[{\citenamefont{Kokorowski and Pritchard}(1997)}]{KoPr:97}
\bibinfo{author}{\bibfnamefont{D.~A.} \bibnamefont{Kokorowski}}
  \bibnamefont{and} \bibinfo{author}{\bibfnamefont{D.~E.}
  \bibnamefont{Pritchard}}, \bibinfo{journal}{J. Mod. Opt.}
  \textbf{\bibinfo{volume}{44}}, \bibinfo{pages}{2575} (\bibinfo{year}{1997}).

\bibitem[{\citenamefont{Dhirani et~al.}(1997)\citenamefont{Dhirani, Kokorowski,
  Rubenstein, Hammond, Rohwedder, Smith, Roberts, and
  Pritchard}}]{DhKoRuHaRo:97}
\bibinfo{author}{\bibfnamefont{A.-A.} \bibnamefont{Dhirani}},
  \bibinfo{author}{\bibfnamefont{D.~A.} \bibnamefont{Kokorowski}},
  \bibinfo{author}{\bibfnamefont{R.~A.} \bibnamefont{Rubenstein}},
  \bibinfo{author}{\bibfnamefont{T.~D.} \bibnamefont{Hammond}},
  \bibinfo{author}{\bibfnamefont{B.}~\bibnamefont{Rohwedder}},
  \bibinfo{author}{\bibfnamefont{E.~T.} \bibnamefont{Smith}},
  \bibinfo{author}{\bibfnamefont{A.~D.} \bibnamefont{Roberts}},
  \bibnamefont{and} \bibinfo{author}{\bibfnamefont{D.~E.}
  \bibnamefont{Pritchard}}, \bibinfo{journal}{J. Mod. Opt.}
  \textbf{\bibinfo{volume}{44}}, \bibinfo{pages}{2583} (\bibinfo{year}{1997}).

\bibitem[{\citenamefont{Rubenstein
  et~al.}(1999{\natexlab{a}})\citenamefont{Rubenstein, Dhirani, Kokorowski,
  Roberts, Smith, Smith, Bernstein, Lehner, Gupta, and
  Pritchard}}]{RuDhKoRoSm:99}
\bibinfo{author}{\bibfnamefont{R.~A.} \bibnamefont{Rubenstein}},
  \bibinfo{author}{\bibfnamefont{A.-A.} \bibnamefont{Dhirani}},
  \bibinfo{author}{\bibfnamefont{D.~A.} \bibnamefont{Kokorowski}},
  \bibinfo{author}{\bibfnamefont{T.~D.} \bibnamefont{Roberts}},
  \bibinfo{author}{\bibfnamefont{E.~T.} \bibnamefont{Smith}},
  \bibinfo{author}{\bibfnamefont{W.~W.} \bibnamefont{Smith}},
  \bibinfo{author}{\bibfnamefont{H.~J.} \bibnamefont{Bernstein}},
  \bibinfo{author}{\bibfnamefont{J.}~\bibnamefont{Lehner}},
  \bibinfo{author}{\bibfnamefont{S.}~\bibnamefont{Gupta}}, \bibnamefont{and}
  \bibinfo{author}{\bibfnamefont{D.~E.} \bibnamefont{Pritchard}},
  \bibinfo{journal}{Phys. Rev. Lett.} \textbf{\bibinfo{volume}{82}},
  \bibinfo{pages}{2018} (\bibinfo{year}{1999}{\natexlab{a}}).

\bibitem[{\citenamefont{Pritchard et~al.}(1999)\citenamefont{Pritchard,
  Rubenstein, Dhirani, Kokorowski, Smith, Hammond, and
  Rohwedder}}]{PrRuDhKoSm:99}
\bibinfo{author}{\bibfnamefont{D.~E.} \bibnamefont{Pritchard}},
  \bibinfo{author}{\bibfnamefont{R.~A.} \bibnamefont{Rubenstein}},
  \bibinfo{author}{\bibfnamefont{A.-A.} \bibnamefont{Dhirani}},
  \bibinfo{author}{\bibfnamefont{D.~A.} \bibnamefont{Kokorowski}},
  \bibinfo{author}{\bibfnamefont{E.~T.} \bibnamefont{Smith}},
  \bibinfo{author}{\bibfnamefont{T.~D.} \bibnamefont{Hammond}},
  \bibnamefont{and}
  \bibinfo{author}{\bibfnamefont{B.}~\bibnamefont{Rohwedder}},
  \bibinfo{journal}{Phys. Rev. A} \textbf{\bibinfo{volume}{59}},
  \bibinfo{pages}{4641} (\bibinfo{year}{1999}).

\bibitem[{\citenamefont{Rubenstein
  et~al.}(1999{\natexlab{b}})\citenamefont{Rubenstein, Kokorowski, Dhirani,
  Roberts, Gupta, Lehner, Smith, Smith, Bernstein, and
  Pritchard}}]{RuKoDhRoGu:99}
\bibinfo{author}{\bibfnamefont{R.~A.} \bibnamefont{Rubenstein}},
  \bibinfo{author}{\bibfnamefont{D.~A.} \bibnamefont{Kokorowski}},
  \bibinfo{author}{\bibfnamefont{A.-A.} \bibnamefont{Dhirani}},
  \bibinfo{author}{\bibfnamefont{T.~D.} \bibnamefont{Roberts}},
  \bibinfo{author}{\bibfnamefont{S.}~\bibnamefont{Gupta}},
  \bibinfo{author}{\bibfnamefont{J.}~\bibnamefont{Lehner}},
  \bibinfo{author}{\bibfnamefont{W.~W.} \bibnamefont{Smith}},
  \bibinfo{author}{\bibfnamefont{E.~T.} \bibnamefont{Smith}},
  \bibinfo{author}{\bibfnamefont{H.~J.} \bibnamefont{Bernstein}},
  \bibnamefont{and} \bibinfo{author}{\bibfnamefont{D.~E.}
  \bibnamefont{Pritchard}}, \bibinfo{journal}{Phys. Rev. Lett.}
  \textbf{\bibinfo{volume}{83}}, \bibinfo{pages}{2285}
  (\bibinfo{year}{1999}{\natexlab{b}}).

\bibitem[{\citenamefont{Kokorowski et~al.}(2000)\citenamefont{Kokorowski,
  Roberts, Rubenstein, Smith, and Pritchard}}]{KoRoRuSmPr:00}
\bibinfo{author}{\bibfnamefont{D.~A.} \bibnamefont{Kokorowski}},
  \bibinfo{author}{\bibfnamefont{T.~D.} \bibnamefont{Roberts}},
  \bibinfo{author}{\bibfnamefont{R.~A.} \bibnamefont{Rubenstein}},
  \bibinfo{author}{\bibfnamefont{E.~T.} \bibnamefont{Smith}}, \bibnamefont{and}
  \bibinfo{author}{\bibfnamefont{D.~E.} \bibnamefont{Pritchard}},
  \bibinfo{journal}{Fortschr. Phys.} \textbf{\bibinfo{volume}{48}},
  \bibinfo{pages}{615} (\bibinfo{year}{2000}).

\bibitem[{\citenamefont{Kapale et~al.}(2003)\citenamefont{Kapale, Qamar, and
  Zubairy}}]{KaQaZu:03}
\bibinfo{author}{\bibfnamefont{K.~T.} \bibnamefont{Kapale}},
  \bibinfo{author}{\bibfnamefont{S.}~\bibnamefont{Qamar}}, \bibnamefont{and}
  \bibinfo{author}{\bibfnamefont{M.~S.} \bibnamefont{Zubairy}},
  \bibinfo{journal}{Phys. Rev. A} \textbf{\bibinfo{volume}{67}},
  \bibinfo{pages}{023805} (\bibinfo{year}{2003}).

\bibitem[{\citenamefont{Szriftgiser et~al.}(1996)\citenamefont{Szriftgiser,
  Gu\'ery-Odelin, Arndt, and Dalibard}}]{SzGuArDa:96}
\bibinfo{author}{\bibfnamefont{P.}~\bibnamefont{Szriftgiser}},
  \bibinfo{author}{\bibfnamefont{D.}~\bibnamefont{Gu\'ery-Odelin}},
  \bibinfo{author}{\bibfnamefont{M.}~\bibnamefont{Arndt}}, \bibnamefont{and}
  \bibinfo{author}{\bibfnamefont{J.}~\bibnamefont{Dalibard}},
  \bibinfo{journal}{Phys. Rev. Lett.} \textbf{\bibinfo{volume}{77}},
  \bibinfo{pages}{4} (\bibinfo{year}{1996}).

\bibitem[{\citenamefont{Salomon et~al.}(1987)\citenamefont{Salomon, Dalibard,
  Aspect, Metcalf, and Cohen-Tannoudji}}]{SaDaAsMeCo:87}
\bibinfo{author}{\bibfnamefont{C.}~\bibnamefont{Salomon}},
  \bibinfo{author}{\bibfnamefont{J.}~\bibnamefont{Dalibard}},
  \bibinfo{author}{\bibfnamefont{A.}~\bibnamefont{Aspect}},
  \bibinfo{author}{\bibfnamefont{H.}~\bibnamefont{Metcalf}}, \bibnamefont{and}
  \bibinfo{author}{\bibfnamefont{C.}~\bibnamefont{Cohen-Tannoudji}},
  \bibinfo{journal}{Phys. Rev. Lett.} \textbf{\bibinfo{volume}{59}},
  \bibinfo{pages}{1659} (\bibinfo{year}{1987}).

\bibitem[{\citenamefont{Cook and Hill}(1982)}]{CoHi:82}
\bibinfo{author}{\bibfnamefont{R.~J.} \bibnamefont{Cook}} \bibnamefont{and}
  \bibinfo{author}{\bibfnamefont{R.~K.} \bibnamefont{Hill}},
  \bibinfo{journal}{Opt. Commun.} \textbf{\bibinfo{volume}{43}},
  \bibinfo{pages}{258} (\bibinfo{year}{1982}).

\bibitem[{\citenamefont{Balykin et~al.}(1987)\citenamefont{Balykin, Letokhov,
  Ovchinnikov, and Sidorov}}]{BaLeOvSi:87}
\bibinfo{author}{\bibfnamefont{V.}~\bibnamefont{Balykin}},
  \bibinfo{author}{\bibfnamefont{V.}~\bibnamefont{Letokhov}},
  \bibinfo{author}{\bibfnamefont{Y.~B.} \bibnamefont{Ovchinnikov}},
  \bibnamefont{and} \bibinfo{author}{\bibfnamefont{A.}~\bibnamefont{Sidorov}},
  \bibinfo{journal}{JETP Lett.} \textbf{\bibinfo{volume}{45}},
  \bibinfo{pages}{282} (\bibinfo{year}{1987}).

\bibitem[{\citenamefont{Balykin et~al.}(1988)\citenamefont{Balykin, Letokhov,
  Ovchinnikov, and Sidorov}}]{BaLeOvSi:88}
\bibinfo{author}{\bibfnamefont{V.}~\bibnamefont{Balykin}},
  \bibinfo{author}{\bibfnamefont{V.}~\bibnamefont{Letokhov}},
  \bibinfo{author}{\bibfnamefont{Y.~B.} \bibnamefont{Ovchinnikov}},
  \bibnamefont{and} \bibinfo{author}{\bibfnamefont{A.}~\bibnamefont{Sidorov}},
  \bibinfo{journal}{Phys. Rev. Lett.} \textbf{\bibinfo{volume}{60}},
  \bibinfo{pages}{2137} (\bibinfo{year}{1988}).

\bibitem[{\citenamefont{Kasevich et~al.}(1990)\citenamefont{Kasevich, Weiss,
  and Chu}}]{KaWeCh:90}
\bibinfo{author}{\bibfnamefont{M.~A.} \bibnamefont{Kasevich}},
  \bibinfo{author}{\bibfnamefont{D.~S.} \bibnamefont{Weiss}}, \bibnamefont{and}
  \bibinfo{author}{\bibfnamefont{S.}~\bibnamefont{Chu}}, \bibinfo{journal}{Opt.
  Lett.} \textbf{\bibinfo{volume}{15}}, \bibinfo{pages}{607}
  (\bibinfo{year}{1990}).

\bibitem[{Hil()}]{Hilbert}
\bibinfo{note}{Formally this means $\int \!\d{k'}
  \mathscr{P}\frac{1}{k'-k}\mathscr{P}\frac{1}{k'-k''} = \pi^2\delta(k-k'')$.}

\bibitem[{inf()}]{infty}
\bibinfo{note}{When $\tau, \alpha \to \infty$, then $I \to \delta$ and
  $N_\t{refl} \to 1$, as it should.}

\bibitem[{T()}]{T}
\bibinfo{note}{It may be tempting to take the expression on the l.h.s. of
  Eq.~(\ref{three}), considered as a function of $T$ with $x_\t{M}$ fixed, as
  proportional to the arrival time probability density at the mirror, but this
  differs in general from the quantum expression of \cite{Kijowski-RMP-1974}
  J.~Kijowski, Rep. Math. Phys. {\bf 6},361 (1974), recently discussed e.g. in
  \cite{HaHeMu:05} V.~Hannstein, G.~C.~Hegerfeldt and J.~G.~Muga, J. Phys. B
  {\bf 38}, 409 (2005). For the parameters of Fig. \ref{fig:tempslit_x_exp},
  however, there is good agreement between both.}

\bibitem[{\citenamefont{Kijowski}(1974)}]{Kijowski-RMP-1974}
\bibinfo{author}{\bibfnamefont{J.}~\bibnamefont{Kijowski}},
  \bibinfo{journal}{Rep. Math. Phys.} \textbf{\bibinfo{volume}{6}},
  \bibinfo{pages}{361} (\bibinfo{year}{1974}).

\bibitem[{\citenamefont{Hannstein et~al.}(2005)\citenamefont{Hannstein,
  Hegerfeldt, and Muga}}]{HaHeMu:05}
\bibinfo{author}{\bibfnamefont{V.}~\bibnamefont{Hannstein}},
  \bibinfo{author}{\bibfnamefont{G.~C.} \bibnamefont{Hegerfeldt}},
  \bibnamefont{and} \bibinfo{author}{\bibfnamefont{J.~G.} \bibnamefont{Muga}},
  \bibinfo{journal}{J.~Phys.~B: At.~Mol.~Opt.~Phys.}
  \textbf{\bibinfo{volume}{38}}, \bibinfo{pages}{409} (\bibinfo{year}{2005}).

\end{thebibliography}
\end{document}